\documentclass[prl,twocolumn,showpacs,preprintnumbers,amsmath,amssymb]{revtex4}
\usepackage{graphicx}
\usepackage{dcolumn}
\usepackage{bm}

\begin{document}
\preprint{PREPRINT}

\title{Surface Plasmon Interference Fringes in Back-Reflection}

\author{A.~Drezet, A.~L.~Stepanov, A.~Hohenau,
B.~Steinberger, N.~Galler, H.~Ditlbacher, A.~Leitner,
F.~R.~Aussenegg, and J.~R.~Krenn}
 \affiliation{Institut of physics,
Karl-Franzens University Graz, Universit\"atsplatz 5 A-8010 Graz,
Austria}

\author{M.~U.~Gonzalez, and J.-C.~Weeber}
 \affiliation{Laboratoire de Physique de l'universit\'{e} de Bourgogne, UMR CNRS 5027, 9
 Avenue A.~Savary, BP 47870, F-21078 Dijon, France}

\date{\today}
\begin{abstract}

We report the experimental observation of surface plasmon
polariton (SPP) interference fringes with near-unity visibility
and half-wavelength periodicity obtained in back reflection on a
Bragg mirror. The presented method based on leakage radiation
microscopy
 (LRM) represents an alternative solution to optical near-field analysis
and opens new ways for the quantitative analysis of SPP fringes.
With LRM we investigate various SPP interference patterns and
analyze the high reflectivity of Bragg mirror in comparison with
theoretical models.

\end{abstract}
\pacs{78.~67.-n}
 \maketitle
The development of optics at the micron and sub-micron scales
requires the control over coherence of wave propagation in a
confined environment as well as the knowledge of the optical
properties of the structures used for this purpose. In this
context the recent progresses in surface plasmon polariton
(SPPs)~\cite{Raether} optics allowed the realization of optical
elements as Bragg mirrors and beam splitters. These elements have
been combined in various 2D SPP devices like a Mach-Zehnder
interferometer~\cite{Harry}, or an elliptical resonator
\cite{Drezet}. In order to achieve SPP mirrors with very high
reflectivity one has to study quantitatively the interaction of
SPP waves and Bragg reflectors. In this context the quantitative
near-field measurement of the reflectivity of Bragg mirrors
integrated into SPP waveguides~\cite{Weeber1, Weeber2} was
reported. Here we propose an alternative method of analysis based
on far-field leakage radiation microscopy (LRM) that we developed
recently into a systematic routine \cite{Andrey}. For this purpose
we consider the back-reflection of propagating SPP waves impinging
normally onto a Bragg mirror. In order to show the accuracy of the
method we report experimental evidence for interference between
incident and back-reflected SPP wave with a periodicity of
$\lambda_{\textrm{SPP}}/2$ ( $\lambda_{SPP}$ being the SPP
wavelength) and with a quasi-unity fringe visibility $V\simeq1$.
We show that the results are consistent with the theoretical
expectations and with an ideal SPP reflectivity of 100\%. This in
turn proves that back-reflection can be a useful tool for the
local tailoring of the SPP intensity pattern by interference of
incident and reflected SPP beams.
\begin{figure}[h]
\includegraphics[width=10cm]{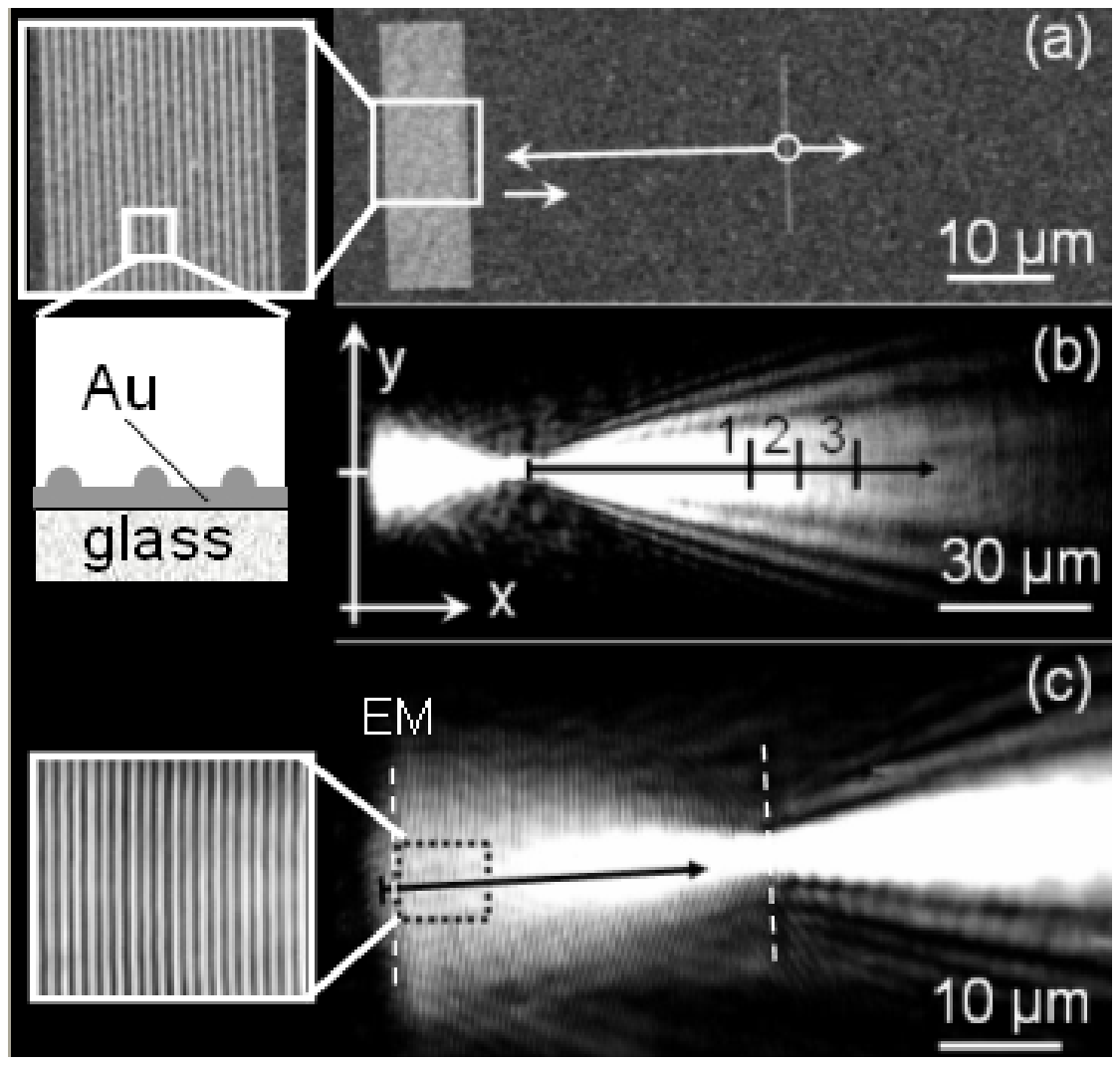}
\caption{(a) Scanning electron microscope (SEM) image of Bragg
mirror and launching ridge. The insets shows a zoom on the Bragg
mirror and a sketch of the mirror structure. SPPs are launched
from the ridge region contained in the white circle and propagate
normally to the ridge as represented by the arrows. The left SPP
beam is back-reflected by the mirror. (b) LRM image of SPP
interference pattern. The black arrow and the marks 1,2,3 refer to
cross-cuts in Fig.~3. (c) LRM image at higher magnification
allowing the observation of interference fringes between mirror
and ridge. The black arrow refers to the cross-cut in Fig.~4. The
inset shows a zoom of the fringes. The white dotted lines show the
position of the ridge and of the effective Brag mirror (EM) (see
Fig.~4).} \label{fig1}
\end{figure}
The Bragg mirror structures are constituted by a set of 20
parallel gold ridges (70 nm height, 150 nm width) fabricated by
electron beam lithography (EBL) on glass subtrate \cite{Weeber1}.
The distance between the ridges is $d=\lambda_{SPP}/2=390$ nm (see
Fig.~1a). At normal incidence this corresponds to Bragg
reflectance maxima for a laser wavelength
$\lambda_{0}\simeq\lambda_{\textrm{SPP}}$ of 780 nm (Ti:sapphire).
SPP waves are locally launched from an additional ridge (150 nm
width) located in front of the Bragg reflector at a distance of
$\simeq30$ $\mu$m (see Fig.~1a). The laser beam is focused onto
this ridge using a microscope objective (50$\times$, numerical
aperture 0.7) and the laser spot diameter is around 2 $\mu$m which
corresponds to a full divergence angle $\theta$ of the SPP beam on
the gold film of about $32^\circ$. The SPP wave propagating to the
left impinges normally on the Bragg mirror and is reflected back
to the right giving rise to an interference pattern. Fig.~1b shows
a LRM image \cite{Andrey, Drezet} of the SPP propagation obtained
using an oil immersion objective (63 $\times$, numerical
aperture=1.25). The absence of SPP transmittance behind the Bragg
mirror leads to the conclusion that the incident SPP wave is
mostly back-reflected.
\begin{figure}[h]
\includegraphics[width=2.5in]{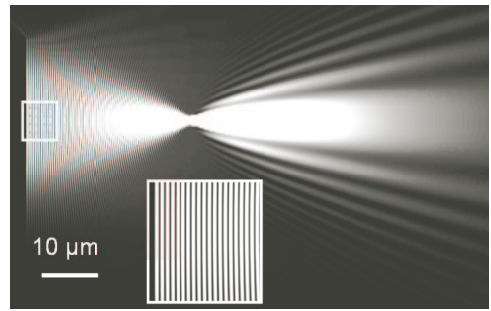}
\caption{Theoretical simulation of the SPP interference pattern in
Fig.~1. The model relies on scalar waves as discussed in the text.
The inset shows the SPP fringes at higher magnification.}
\label{fig2}
\end{figure}
The diffraction behavior visible to the right of the launching
ridge can be straightforwardly understood as a result of the
interference between the SPP wave originating from the ridge and a
virtual image-source located in the half-space to the left of the
Bragg mirror. This is confirmed theoretically by simulating the
SPP propagation with a simple scalar wave model \cite{Drezet} (see
Fig.~2) . The SPP field launched at the ridge is modelled by a
continuous distribution of 2D dipoles whose orientation is in the
sample plane and normal to the ridge direction. To simplify the
calculation we considered a SPP mirror consisting of a single
ridge located at the position of the Bragg reflector. The
interaction of the incident SPP beam with this ideal mirror
generates secondary SPP waves interfering with the incident beam
to generate the observed lateral interference patterns to the
right of the ridge. It is observed that despite the high value of
the divergence angle $\theta$ the observed reflectivity is very
high. This shows qualitatively that the reflectivity of this Bragg
mirror is not very sensible to the incidence angle $\alpha$ in a
domain $\Delta \alpha\simeq \pm 10^\circ$. This effect has already
been observed in the near-field experiments mentioned
\cite{Weeber1} and can be explained if we suppose that a
periodical grating opens a gap in the angular dispersion curve of
SPPs \cite{Barnes}.
\begin{figure}[h]
\includegraphics[width=9 cm]{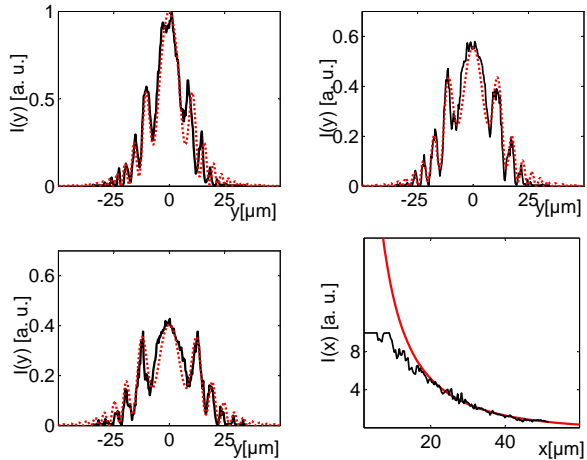}
\caption{(color online) (a-c) Transversal cross-cuts of the SPP
intensity corresponding to the marks 1,2,and 3 in Fig.~1b
respectively. The experimental data (black curves) and the model
(red dashed curves) are compared. The origin of the y axis is
given by a white mark on the y axis of Fig.~1b (d) Longitudinal
cross-cuts along the direction indicated by an arrow in Fig.~1b.
The origin of the arrow corresponds to the origin of the
cross-cut. Experimental data (black curve) and theory (red curve)
are compared.} \label{fig3}
\end{figure}
In order to analyze the SPP propagation we extracted different
cross-cuts of the LRM images. The three transversal cross-cuts
represented in Figs.~3a, b and c correspond to the directions and
positions indicated in Fig.~1b. Fig.~3d shows a longitudinal
cross-cut along the SPP propagation direction in Fig.~1b.
Altogether these cross-cuts agree very well with the simulation if
we adjust adequately the SPP propagation length $L_{\textrm{SPP}}$
and the effective mirror reflectivity $R$ which are the two free
parameters of the model. Both transversal and longitudinal
cross-cuts are necessary for fixing $R$ and $L_{\textrm{SPP}}$. We
find the best fit with a total reflectivity $R\simeq 95\pm5$\% and
$L_{\textrm{SPP}}\simeq 26 \mu$m which agrees with the value
expected for a gold film \cite{Raether}.
\begin{figure}[h]
\includegraphics[width=9cm ]{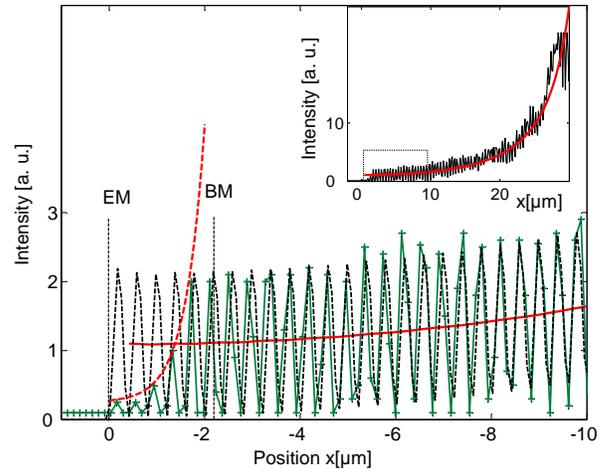}
\caption{(color online) Cross-cut of the interference fringes
(green curve) along the direction represented by the black arrow
of Fig.~1c (the origin of the cross-cut axis corresponds to the
origin of the arrow). The red curve shows the averaged intensity
as predicted by the model and  the black dashed curve shows the
predicted interference. The red dashed curve is the exponential
fit for the effective plasmon damping starting at the beginning of
Bragg's mirror (BM) and finishing at the position of the effective
mirror (EM). The inset shows a larger view of the interference
(black curve) as well as the predicted averaged intensity (red
curve).} \label{fig4}
\end{figure}
To understand the physical meaning of the $R$ value we now focus
our attention on the remarkable interference features present in
the region between the Bragg reflector and the launching ridge.
Indeed, in this region, as shown in the simulation (see inset on
Fig.~2), one expects theoretically interferences fringes with very
high contrast
$V=(I_{\textrm{max}}-I_{\textrm{min}})/(I_{\textrm{max}}+I_{\textrm{min}})$
close to one. We note that this effect is not visible in Fig.~1b,
since this figure is resolution limited due to the finite size of
the pixels on the charged-coupled-device camera used to acquire
the image. By increasing the magnification of the LRM \cite{note2}
we however overcome this limit and we observe fringe interference
with a periodicity of 390 nm corresponding to
$\lambda_{\textrm{SPP}}/2$ (see Fig.~1c). This observation can
seem rather surprising if we think of Rayleigh diffraction
limiting the spatial resolution to
$\sim\lambda_{0}/(2NA)\simeq\lambda_{\textrm{SPP}}/2.5$ with
$NA=1.25$, the oil immersion objective numerical aperture. One
should intuitively thus expect that the fringes will be blurred in
the image plane, as indeed incoherent optics predicts a reduced
visibility below $10$\%  \cite{note}. However, the SPP standing
wave pattern is a coherent phenomenon and it can be proven using
Fourier optics and Abbe's theory \cite{Teich} that the diffraction
limit does not affect the visibility $V$\cite{Teich2}. In the
present case the immersion objective has a high numerical aperture
and we estimate that the unity fringe visibility existing in the
sample plane conserves its value in the image plane where we
expect thus
$V_{\textrm{exp}}\simeq 1$.\\
In order to analyze the fringes we plot a cross-cut in the
direction indicated by the black arrow in Fig.~1c. This cross cut
is depicted in the inset of Fig.~4. The red curve shows the
averaged intensity as simulated by the theoretical model which
reproduces the general trend of the intensity profile quite well.
The agreement is however not limited to the averaged intensity as
visible from the main curve of Fig.~4 which zooms a detail of the
inset close to the Bragg mirror. The oscillation fringes observed
experimentally coincide with the theoretical predictions and in
particular reproduce the very high contrast expected in this
region. A simple way to calculate the reflectivity $R$ is to
consider the ideal standing wave pattern obtained using two
counter propagative plane waves leading to the intensity
$I(x)\simeq 1+V_{\textrm{ideal}}\cos{(4\pi
x/\lambda_{\textrm{SPP}}+const.)}$ with
$V_{\textrm{ideal}}=2\sqrt{(R)}/(1+R)$. Since this plane wave
hypothesis can be considered as valid in the vicinity of the Bragg
mirror boundary we deduce that $R$ should equal $95\pm 5$\%. The
uncertainty is partly a result of the microscope diffraction limit
discussed before. It can be added that the consistency of the
argumentation is experimentally confirmed by the fact that an
incoherent illumination of the sample with an external light
source (not shown) does not allow us to resolve the separation
between the ridges constituting the Bragg mirror. Since the
periodicity of the fringes equals the one of the Bragg mirror it
indeed proves that an incoherent effect can not be invoked as a
justification for the quasi-unity fringes visibility observed with
SPPs. From these results and from the experimental fact that no
SPP is transmitted through the Bragg mirror (see Figs.~1 and 4)
one can thus conclude that 1) the transmission $T$ is below the
noise level, i.~e., below 0.5\% and 2) that a few percent of the
the incident SPP beam (i.~e., S=5-10\%) are scattered to light
into the glass substrate, this to accord to energy conservation.
It can be finally remarked that the mirror model used in this work
did not consider, to simplify the problem, the precise structure
of the Bragg mirror. This explains why in order to reproduce
accurately the experimental observation we must position precisely
the effective mirror (EM in Fig.~4) 2 $\mu$m to the left of the
physical boundary corresponding to the first ridge of the Bragg
mirror (BM in Fig.~4). The experimentally observed effective
damping inside of Bragg's mirror can be physically understood if
we consider that a SPP wave needs to propagate inside of the
periodical structure before being reflected back. This damping can
be well approximated by an exponential decay law whose propagation
length $\delta$ is around 500 nm.

In summary, we have presented an accurate method to measure
reflectivity for SPP Bragg mirrors in normal incidence. The method
based on LRM imaging allows the observation of standing wave
fringes with a periodicity as small as $\lambda_{\textrm{SPP}}/2$
and with a high contrast close to unity. The method due its
sensibility constitutes a good complement to near-field
measurements made on the same kind of structures \cite{Weeber2}.
In all cases the observations are in good quantitative agreement
with the theoretical expectations and show that the essential
behaviors are well understood.

For financial support the European Union, under projects FP6
NMP4-CT-2003-505699 and the Lisa Meitner programm of the Austrian
Science Foundation (M868-N08) are acknowledged.


\begin{thebibliography}{99}
\bibitem{note2}
By increasing by a factor 5 the focal length of the imaging lens
located in front of the CCD camera (see \cite{Andrey}) we multiply
the microscope magnification by the same factor.
\bibitem{note}
From incoherent optics the original intensity pattern
$I_{0}(\mathbf{x})$ (where $\mathbf{x}=[x,y]$ is the 2D position
vector) is convoluted with the impulse response function of the
objective $F^{2}(\mathbf{x})=[2J_{1}(k|\mathbf{x}|
NA)/(k|\mathbf{x}| NA)]^2$. For the intensity pattern in the image
plane one has $I_{i}(\mathbf{x})\propto\int F^{2}(\mathbf{u})\cdot
I_{0}(\mathbf{u}-\mathbf{x}/M)d^{2}\mathbf{u}$ with $M$ the
absolute value of the transverse microscope magnification. If
$I_{0}(\mathbf{x})=1+\cos{(2\mathbf{k}\cdot\mathbf{x})}$ one
obtains $I_{i}(\mathbf{x})\propto
1+V\cos{(2\mathbf{k}\cdot\mathbf{x}/M)}$ with $V\simeq
e^{2\cdot(1.4/NA)^2}\simeq 9$\% (we used the numerical
approximation $[2J_{1}(u)/(u)]^2\approx
e^{-\frac{1}{2}(u/1.4)^2}$).
\bibitem{Teich2}
For the electric field of a coherent SPP wave which in the object
plane has the form
$f_{0}(\mathbf{x})=e^{i\mathbf{k}\cdot\mathbf{x}}+\mathcal{R}\cdot
e^{-i\mathbf{k}\cdot\mathbf{x}}$ ($R=|\mathcal{R}|^{2}$ is the
Bragg mirror reflectivity) one obtains in the image plane
$f_{i}(\mathbf{x})\propto\int F(\mathbf{u})\cdot
f_{0}(\mathbf{u}-\mathbf{x}/M)d^{2}\mathbf{u}$, e.~g.~, the
convolution of $f_{0}$ with the impulse response of the objective
for the field $F(\mathbf{x})=2J_{1}(k|\mathbf{x}|
NA)/(k|\mathbf{x}| NA)$(compare \cite{note}). Since
$F(\mathbf{x})=F(-\mathbf{x})$ one deduces
$f_{i}(\mathbf{x})\propto
e^{-i\mathbf{k}\cdot\mathbf{x}/M}+\mathcal{R}\cdot
e^{i\mathbf{k}\cdot\mathbf{x}/M}$. It implies
$I_{i}(\mathbf{x})=|f_{i}(\mathbf{x})|^{2}\propto
1+V\cos{(2\mathbf{k}\cdot\mathbf{x}/M+const.)}$ with the same
visibility $V=2\sqrt{(R)}/(1+R)$ than the SPP wave intensity
$I_{0}(\mathbf{x})=|f_{0}(\mathbf{x})|^{2}\propto
1+V\cos{(2\mathbf{k}\cdot\mathbf{x}-const.)}$.
\end{thebibliography}
\end{document}